\documentclass{emulateapj}
\usepackage{amsmath}
\usepackage{natbib}
\usepackage{multirow}
\newcommand{\gps}{\ensuremath{g_{\rm P1}}}

\newcommand{\ips}{\ensuremath{i_{\rm P1}}}
\newcommand{\zps}{\ensuremath{z_{\rm P1}}}
\newcommand{\yps}{\ensuremath{y_{\rm P1}}}

\newcommand{\PS}{\protect \hbox {Pan-STARRS1}}
\newcommand{\degree}{\ensuremath{^\circ}}
\newcommand{\dfplot}[1]{\plotone{#1}}

\newcommand{\m}{\ensuremath{\vec{m}}}
\renewcommand{\mp}{\ensuremath{\left\{ \m \right\}}}
\renewcommand{\a}{\ensuremath{\vec{\alpha}}}

\newcommand{\kms}{\ensuremath{\mathrm{km}~\mathrm{s}^{-1}}}

\newcommand{\HI}{H\,{\scriptsize I}}
\newcommand{\HII}{H\,{\scriptsize II}}
\newcommand{\Ha}{\ensuremath{\mathrm{H\alpha}}}

\shorttitle{3D Dust Mapping the Orion Molecular Complex}
\shortauthors{E. F. Schlafly et al.}
\begin{document}
\title{3D Dust Mapping Reveals that Orion Forms Part of a Large Ring of Dust}
\author{
E. F. Schlafly,\altaffilmark{1}
G. Green,\altaffilmark{2}
D. P. Finkbeiner,\altaffilmark{2,3}
H.-W. Rix,\altaffilmark{1}
W. S. Burgett,\altaffilmark{4}
K. C. Chambers,\altaffilmark{4}
P. W. Draper,\altaffilmark{5}
N. Kaiser,\altaffilmark{4}
N. F. Martin,\altaffilmark{6,1}
N. Metcalfe,\altaffilmark{5}
J. S. Morgan,\altaffilmark{4}
P. A. Price,\altaffilmark{7}
J. L. Tonry,\altaffilmark{4}
R. J. Wainscoat,\altaffilmark{4}
C. Waters\altaffilmark{4}
}

\altaffiltext{1}{Max Planck Institute for Astronomy, K\"{o}nigstuhl 17, D-69117 Heidelberg, Germany}
\altaffiltext{2}{Harvard-Smithsonian Center for Astrophysics, 60 Garden Street, Cambridge, MA 02138}
\altaffiltext{3}{Department of Physics, Harvard University, 17 Oxford Street, Cambridge MA 02138}
\altaffiltext{4}{Institute for Astronomy, University of Hawaii, 2680 Woodlawn Drive, Honolulu HI 96822}
\altaffiltext{5}{Department of Physics, Durham University, South Road, Durham DH1 3LE, UK} 
\altaffiltext{6}{Observatoire Astronomique de Strasbourg, CNRS, UMR 7550, 11 rue de l'Universit\'{e}, F-67000 Strasbourg, France}
\altaffiltext{7}{Department of Astrophysical Sciences, Princeton University, Princeton, NJ 08544, USA} 

\begin{abstract}
The Orion Molecular Complex is the nearest site of ongoing high-mass star formation, making it one of the most extensively studied molecular complexes in the Galaxy.  We have developed a new technique for mapping the 3D distribution of dust in the Galaxy using Pan-STARRS1 photometry.  We isolate the dust at the distance to Orion using this technique, revealing a large (100~pc, 14\degree\ diameter), previously unrecognized ring of dust, which we term the ``Orion dust ring.''  The ring includes Orion A and B, and is not coincident with current \Ha\ features.  The circular morphology suggests formation as an ancient bubble in the interstellar medium, though we have not been able to conclusively identify the source of the bubble.  This hint at the history of Orion may have important consequences for models of high-mass star formation and triggered star formation.
\end{abstract}


\section{Introduction}
\label{sec:intro}

The Orion Molecular Complex is the nearest site of active high-mass star formation, and is consequently among the most extensively studied regions in the Galaxy \citep{Bally:2008b}.  The overall objective of these studies has been to understand the relationship between stars and gas: the formation of stars from molecular clouds, the destruction of the clouds by newly formed stars, and the cooling of the gas to form new molecular clouds.  In this work we present a new, three-dimensional dust map of the Orion Molecular Complex that reveals a large ring of dust, of which the main Orion molecular clouds form a part.

The circular morphology of the ring suggests that it was formed as a bubble in the interstellar medium (ISM).  Bubble structures are common in the ISM over a range of scales, from small bubbles around planetary nebulae, to larger \HII\ regions around young stars, to still larger supernova-driven bubbles, and finally to huge superbubbles formed by clusters of young stars and their associated supernovae \citep{Koo:1992}.  These bubbles violently reshape the ISM, potentially triggering star formation in some places while extinguishing it in others, compressing the ISM to form new clouds but blowing apart the original clouds.  These bubbles are common, filling a significant fraction of the Galaxy's volume with the hot, ionized phase of the insterstellar medium \citep{McKee:1995, Ferriere:2001}. 

Accordingly, examples of bubbles throughout the Galaxy should be common.  Indeed, the sun resides in the aptly named Local Bubble, which has recently been mapped in exquisite detail by \citet{Lallement:2014}.  The Orion Molecular Complex is already home to another of the best studied superbubbles: the Orion-Eridanus superbubble \citep{Heiles:1976, Reynolds:1979}.  On smaller scales, bubble features in the ISM are numerous: the Milky Way Project, for instance, has found thousands of bubbles in infrared light observed by {\it Spitzer} \citep{Kendrew:2012, Simpson:2012}.

In this work, we focus on a region of sky surrounding the Orion Molecular Complex.  This region is already known to host a number of bubble-like structures---unsurprisingly, given the history of high mass star formation in the area.  These features include the spectacular parsec-scale Orion Nebula (see \citealt{ODell:2001} for a review) and its associated x-ray bubble \citep{Guedel:2008}, the 30~pc radius molecular ring surrounding $\lambda$~Orionis \citep{Maddalena:1987}, and Barnard's Loop, a striking 60~pc radius half ring seen in \Ha\ \citep{Barnard:1894}.  The Orion-Eridanus superbubble is the largest bubble in the region, and is up to 300~pc in size.

Gas and dust are often good tracers of these bubble structures, most dramatically illustrated by the ring of dust and molecular gas in the $\lambda$~Orionis molecular ring.  Traditional two-dimensional dust maps are limited in their usefulness for tracing bubble structures in Orion, however, due to contamination with other dust structures along the line of sight.  The region surrounding Orion, which lies at a distance of about 400~pc, is also home to the Monoceros R2 molecular cloud at 830~pc.  The Orion Northern Filament, meanwhile, extends close to the Galactic plane, where much dust at greater distances is present.

We overcome these limitations by mapping dust in the vicinity of Orion in three dimensions.  We use the technique of \citet{Green:2014}, taking advantage of high-quality optical photometry from \PS\ to find the distances and reddenings to stars.  This effort is related to that of \citet{Lallement:2014}, \citet{Sale:2014}, \citet{Chen:2014}, \citet{Hanson:2014}, and \citet{Marshall:2006}, who also map the Galaxy's dust in 3D.  The work of \citet{Schlafly:2014} and \citet{Schlafly:2014b} has recently used data from \PS\ to create a catalog of distances to major dust clouds and to map the angular structure of the dust, demonstrating the value of this technique.  In this work, we use the 3D analysis to exclude dust foreground and background to Orion from the region, revealing a large, 100~pc ring of dust that encompasses the main Orion molecular clouds, which we term the ``Orion dust ring.''

The outline of this paper is as follows.  In \textsection \ref{sec:data}, we describe the \PS\ data we use to study the dust near Orion.  We briefly explain our technique for separating the dust in Orion from more nearby and more distant dust in \textsection \ref{sec:method}.  We lay out our observational results in \textsection \ref{sec:results}, and then discuss the implications of these results for the larger Orion Molecular Complex in \textsection \ref{sec:discussion}.  Finally, we conclude in \textsection \ref{sec:conclusion}.

\section{Pan-STARRS1}
\label{sec:data}

Our work uses stellar photometry from \PS\ to study the dust.  The \PS\ imager is composed of a 1.8~m telescope on Haleakala on Maui.  The imager is optimized for survey operations, featuring a wide 3\degree\ field of view outfitted with the billion-pixel GPC1 camera \citep{PS1_optics, PS1_GPCA, PS1_GPCB}.  This work relies on data from the \PS\ $3\pi$ survey, which is designed to observe three-quarters of the sky four times a year, in each of 5 filters \citep{Magnier:2013}.  These filters include the SDSS-like $gr\ips$ as well as \zps\ and \yps, at which wavelengths the \PS\ system is much more sensitive than the SDSS \citep{PS_lasercal}.  The single-epoch depth of the $3\pi$ survey is about 22.0, 22.0, 21.9, 21.0 and 19.8 in $griz\yps$, with stacked depths expected to reach 1.1 magnitudes deeper  \citep{Metcalfe:2013}.  The survey is automatically processed by the \PS\ Image Processing Pipeline, which performs dark, bias, and fringe correction; astrometry, stellar photometry, and galaxy photometry; image stacking and differencing; and catalog construction  \citep{PS1_IPP, PS1_photometry, PS1_astrometry}.  The photometric calibration of the survey, both absolute \citep{JTphoto} and relative \citep{Schlafly:2012}, is better than 1\%.

We use data from ``processing version 1'' of the PS1 data, the first uniform reprocessing of the PS1 data, which is composed primarily of data taken between May 2010 and March 2013.  We use only observations of stars observed in the \gps\ band and at least three other PS1 bands in photometric conditions, which include the entire region around Orion studied here.  Only scattered small areas of sky lack a single band of PS1 coverage.  We have found from repeat observations of point sources that the PS1 pipeline underestimates the photometric uncertainty $\sigma$ it reports for sources, and accordingly use $\sigma = \sqrt{(1.3\cdot\sigma_\mathrm{PS1})^2+0.015^2}$.  

The vast majority of objects in this low-Galactic latitude region are expected to be stars.  Nevertheless, we exclude galaxies by requiring that the aperture magnitudes be brighter than the PSF magnitudes of sources by less than 0.1~mag in at least three bands.  Comparison with somewhat deeper SDSS imaging suggests that this leaves about 12 galaxies per square degree in our catalog, or about 0.4\% of the sources near the north Galactic pole and a much smaller fraction of the sources in the Orion Molecular Complex.

The requirement that stars be detected in \gps\ means that highly reddened stars will not be detected in the survey.  We find in \citet{Schlafly:2014b} that our reddening map begins to saturate at about 1.5~mag $E(B-V)$.  Most of the area considered in this work has reddening smaller than this, but in the dense regions of Orion A and B the map is inaccurate.

\section{Method}
\label{sec:method}

We map the 3D structure of the dust in the vicinity of Orion using stellar photometry from \PS.  We use the method of \citet{Green:2014} to estimate the reddenings and distances to individual stars from their \PS\ photometry.  The resulting distance and reddening uncertainties are highly covariant and non-Gaussian, requiring that we track the full probability distribution function (PDF) of reddening and distance.  We then combine these reddening and distance estimates together for all of the stars along $7^\prime \times 7^\prime$ lines of sight, to estimate the reddening profile $E(D)$, that is, the dust column as a function of distance.  We briefly summarize the method for estimating reddenings to individual stars, and then for estimating reddening profiles.  For details see \citet{Green:2014}.

We compare the observed photometry of stars in the PS1 bands with model photometry of stars with particular temperatures, metallicities, distances, and reddenings.  We assign each model star a probability based on the agreement between its model photometry and the observed photometry and associated uncertainties.  By marginalizing over the model metallicities and temperatures, we obtain PDFs $p(E,D)$ describing the estimated reddenings and distances to stars.

We construct our model for the intrinsic, unreddened colors of stars by fitting a spline to the observed stellar locus of bright stars near the north Galactic pole, which we deredden following the \citet{Schlegel:1998} far-infrared dust map ($E(B-V) \approx 15~\mathrm{mmag}$).  We assume that the PS1 colors of stars depend only on temperature and are independent of metallicity, and adopt the results of \citet{Ivezic:2008} and \citet{Juric:2008} to obtain the absolute magnitudes corresponding to stars of different intrinsic colors as a function of metallicity.  We supplement the main-sequence results of \citet{Ivezic:2008} with analogous relations for giants (Z. Ivezi\'c, private communication).  We assume that the $R_V=3.1$ reddening law of \citet{Fitzpatrick:1999}, adapted to the \PS\ bands by \citet{Schlafly:2011}, describes the effect of dust reddening on our data.  We use a fixed reddening vector in this analysis; in principle, the reddening vector varies depending on spectral type and the total amount of reddening, but only at the several percent level at reddenings of about one magnitude.

To find the probability of a model given observed photometry, we include priors based on the number densities of stars of particular types in the Galaxy.  We use a luminosity function from \citet{Bressan:2012}, the Galactic model for stellar number density of \citet{Juric:2008}, and the Galactic metallicity distribution of \citet{Ivezic:2008}.  Because we assume that the PS1 colors are independent of metallicity, the metallicities we infer are entirely dependent on the Galactic metallicity distribution prior we adopt.  These metallicities in turn affect our derived absolute magnitudes and distances.  We multiply these priors by a Gaussian likelihood function, giving our final posterior PDF.  The likelihood function is a product of Gaussians in the difference between the model magnitudes and the observed magnitudes, with standard deviation equal to the uncertainty in the PS1 magnitudes.  We map out the PDF using an MCMC technique developed by \citet{Goodman:2010} and spelled out by \citet{Foreman-Mackey:2013}, as detailed in \citet{Green:2014}.

Given reddening and distance PDFs for each star, we infer the reddening profile $E(D)$ on each $7^\prime \times 7^\prime$ line of sight.  We parameterize $E(D)$ as a piecewise linear function in the distance modulus $\mu = 5 \log D / (10~\mathrm{pc})$, with separations between adjacent points of $\Delta \mu = 0.53$.  If we label the parameters of this piecewise function $\a$, the probability of a certain set of $\a$ given observed photometry \mp\ is
\begin{equation}
\label{eq:palpha}
p(\a \mid \mp) \propto p(\a) \prod_i \int dD \, p(E(D), D \mid \m_i)
\end{equation}
where $i$ labels different stars, and $p(\a)$ is the prior on $\a$.  We adopt a flat prior on the amount of reddening per bin in $\mu$ for $\a$.  We then perform a nonlinear optimization to find the $\a$ that maximize $p(\a \mid \mp)$, giving us the maximum likelihood reddening profile in each line of sight.  Approximate uncertainties are estimated by finding, at each distance, the range of $E(B-V)$ for which $\chi^2$ increases by one, holding the reddening at other distances constant when possible, given the constraint that reddening must increase with distance.

\begin{figure}[htb]
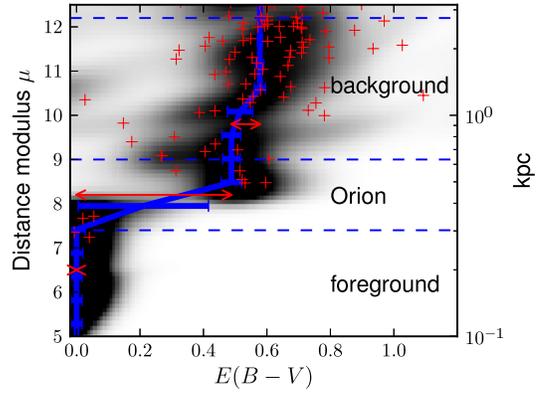

\dfplot{samplos}
\figcaption{
\label{fig:samplos}
The 3D reddening map fit for a single example pixel at $(l, b) = (209.62\degree, -14.17\degree)$.  The grayscale shows the summed probability density functions for all stars along this line of sight, normalized so that the same amount of ink falls in each distance bin.  The red crosses show the expectation values of the reddening and distance modulus for each star.  The blue line shows the inferred maximum likelihood reddening profile and its associated uncertainty.  In this work, we study the dust at the distance to Orion, differentiating it from the dust in its foreground and background; the divisions between these regions are shown with the horizontal dashed lines at 300~pc and 640~pc.  
}
\end{figure}

We use the HEALPix pixelization to define pixels which we fit independently \citep{Gorski:2005}.  We choose a pixel scale of $7^\prime \times 7^\prime$ (HEALPix $N_\mathrm{side} = 512$).  We show one example line of sight in Figure~\ref{fig:samplos}.  The background grayscale shows the summed probability distribution functions for all of the stars on this line of sight, normalized at each distance.  The red crosses show the expectation values of the reddening and distance for each of the stars.  The maximum likelihood for the reddening as a function of distance is shown by the solid blue line.  In this work, we focus on the dust at the distance to Orion as separate from dust foreground and background to the cloud; this separation is shown by the dashed horizontal lines.  There is essentially no foreground dust along this sample line of sight, while there is a clear reddening front at the distance to Orion of about $0.5$~mag $E(B-V)$.  Background to the cloud is an additional approximately $0.1$~mag $E(B-V)$.  These dust columns are shown as red arrows.

The choice of $7^\prime \times 7^\prime$ resolution balances high angular resolution with an adequate number of stars per pixel for obtaining good distance and reddening estimates.  Experiments with lower angular resolutions ($14^\prime$) begin to wash out important features, though we analyze a $14^\prime$ map with improved distance resolution in \textsection\ref{subsec:distance}.  A more principled approach would be to allow the data itself to constrain the resolution of the map, as explored in \citet{Sale:2014b}, or, relatedly, to use a kernel to regularize the map continuously without the need for pixels \citep{Vergely:2001, Vergely:2010, Lallement:2014}.

\section{Results}
\label{sec:results}

\begin{figure*}[htb]
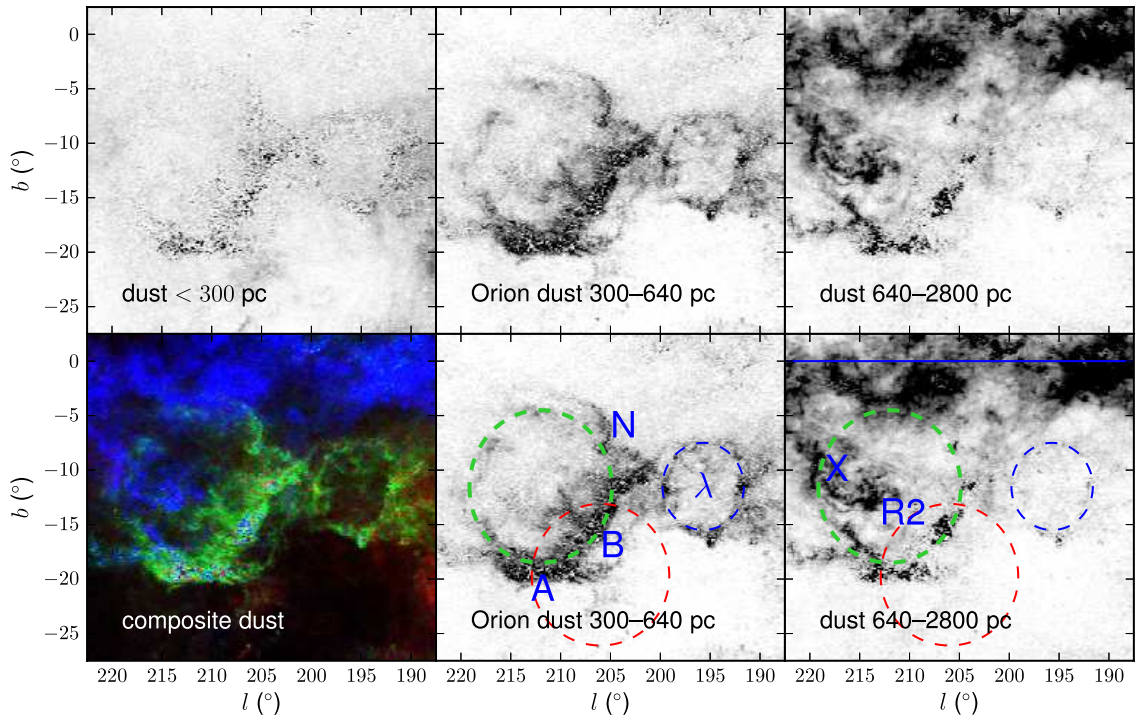

\dfplot{oriondust}
\figcaption{
\label{fig:oriondust}
The 3D distribution of dust towards the Orion Molecular Complex.  The first 3 panels show the column density of dust with distance $<300~\mathrm{pc}$, 300--640~pc, and 640--2800~pc, respectively.  The fourth panel (bottom left) shows a 3-color composite image of these three slices, illustrating the 3D distribution of dust in the region.  Finally, the fifth and sixth panels again show the Orion and more distant dust, this time overplotting circles tracing the various bubble-like structures in the region.  The green dashed circle shows the Orion dust ring; the blue dashed circle shows the $\lambda$~Orionis molecular ring; and the red dashed circle approximately aligns with Barnard's Loop (see Figure~\ref{fig:orionha}).  The last two panels also label the Orion A (A) and Orion B (B) molecular clouds, the Northern Filament (N), the star $\lambda$~Orionis ($\lambda$), Monoceros R2 (R2), the Crossbones (X), and the Galactic plane (horizontal line).  Uncertainty in our distance estimates leaves a faint shadow of Orion in the nearby distance slice.  Differential extinction and an insufficient number of well-observed stars lead to artifacts in the far distance slice through particularly dense clouds in Orion A and B.  White to black corresponds to 0--0.7~mag~$E(B-V)$.  This same scale is used for each of the color planes in the lower left panel.
}
\end{figure*}

On the basis of the above reddening profiles, we produce a 3D map of the dust around Orion, $222.5\degree > l > 187.5\degree$, $-27.5\degree < b < 2.5\degree$.  Within 1~kpc, there are three major structures off the Galactic plane in this field: dust at $\sim 150$~pc associated with Taurus-Perseus-Auriga; dust at $\sim 450$~pc associated with Orion; and dust at $\sim 900$~pc associated with Mon~R2 and the Crossbones \citep{Schlafly:2014}.  To highlight these features we show three slices from our maps in Figure~\ref{fig:oriondust}: the dust with $D<300$~pc, $300<D<640$~pc, and $640<D<2800$~pc.  In each case, the map saturates at 0.7~mag~$E(B-V)$ (2.2~mag~$A_V$).  We also provide a three color composite of these slices, to show how the structures in the various slices line up with one another.  Finally, we show the Orion and more distant dust slices again, overplotting circles that trace the various ring-like features in the region and labelling some of the major dust clouds in the region.  We show the locations of the Orion A (A) and Orion B (B) molecular clouds, as well as the Northern Filament (N) and the location of the star $\lambda$~Orionis ($\lambda$) in the lower middle panel, and the location of the more distant Monoceros R2 (R2), Crossbones (X), and the Galactic plane (horizontal line) in the lower right panel.  We adopt the names and approximate locations of these clouds from \citet{Wilson:2005}.

The second panel of Figure~\ref{fig:oriondust} clearly reveals the ``Orion dust ring,'' a large, 14\degree\ diameter ring of dust at the distance to Orion, centered at $(l, b) = (212\degree, -11.5\degree)$.  This ring circles through Orion A and B, proceeds up the Northern Filament, around foreground to the plane of the Milky Way at $b = -4.8\degree$, and then in front of the Crossbones until rejoining Orion A.  Except for a small patch toward the northeast of the ring, the ring is complete---though the southwestern half of the ring contains vastly more dust than the northeastern.  The Orion dust ring has not been recognized and documented in the literature before.  Adjacent to the ring is a second, smaller, 8\degree\ diameter ring of dust that is the well known molecular ring surrounding $\lambda$~Orionis.  When confusing foreground and background dust is removed, these two rings appear qualitatively equally prominent.

The dust foreground to Orion is concentrated to the south and west of the region.  South of Orion a few foreground filamentary clouds are present, but in general directions toward the ring are free of significant foreground dust.  A faint shadow of Orion A and B is visible in the nearby distance slice owing to the distance uncertainty in the analysis.

The dust background to Orion is complex.  The Mon~R2 cloud projects near the center of the Orion dust ring.  Moreover, filaments in the Crossbones project near the ring edge.  Some material near the eastern edge of Orion A appears to reside near the distance of Mon~R2, making this region highly confused.  In the Galactic plane, a wealth of material at Mon~R2's distance or further is prominent, though with morphology quite different from the structure at Orion's distance. The Crossbones has a filamentary morphology very similar to the dust ring, lining up well in position and angle with the Orion dust ring.  Finally, dense clouds in Orion bleed through into our background dust maps, producing artifacts behind Orion A and B.

The primary cause of these artifacts is significant differential extinction within each pixel and fewer detected background stars.  The work of \citet{Sale:2014} better handles unresolved differential extinction, but we have not pursued that approach here.  This problem leads to artifacts in the background of dense clouds, but is less of an issue in their foreground.  Behind the dense clouds, relatively few stars are detected due to the significant extinction, and these stars have inconsistent reddenings, due to the differential extinction. In the foreground of the dense clouds, on the other hand, the observed stars are largely unextinguished and their reddenings are consistent to within a few hundredths, making the dust mapping straightforward.  For this reason, we are less concerned with background dust contaminating our Orion distance map than of the reverse, though some limited contamination is expected because of uncertainty in the distances.  Consistent with this, the densest parts of Mon~R2 and the Crossbones show very little signature in our Orion distance slice in Figure~\ref{fig:oriondust}.

While we do not expect contamination between the background dust and the Orion dust, the vaguely similar morphology of the Orion filaments in the foreground of the Crossbones and the Crossbones is surprising.  However, most of the Orion dust ring near the Crossbones is interior to the green ring shown in the fifth panel of Figure~\ref{fig:oriondust}, while most of the dust in the Crossbones is just outside this ring.  In detail the fifth and sixth panels of Figure~\ref{fig:oriondust} show that the ring of dust in Orion is poorly correlated spatially with the more distant dust: the ring morphology is not generated by contamination between our different slices.

\begin{figure}[htb]
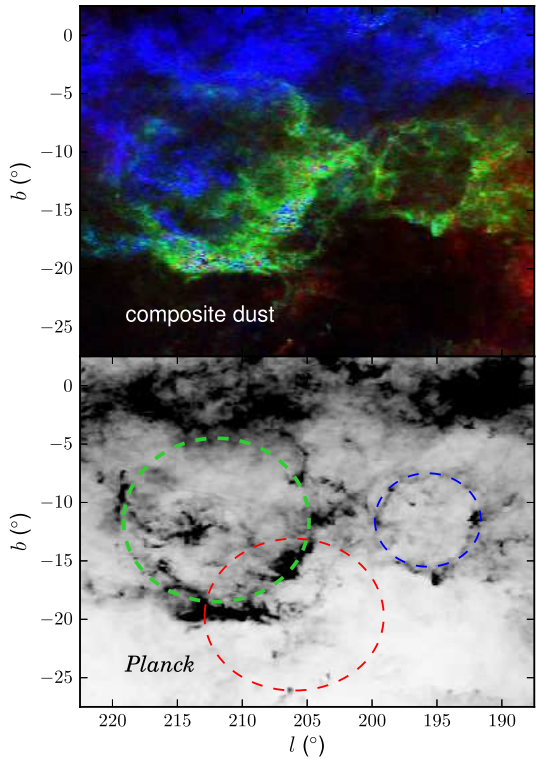

\dfplot{orionplanck}
\figcaption{
\label{fig:orionplanck}
Our 3D dust map of the Orion Molecular Complex (top) compared with the {\it Planck} thermal dust map of the area (bottom).  The color scale is 0--0.7~mag~$E(B-V)$ in each color plane in the top panel, and 0--1.6~mag $E(B-V)$ in the lower panel.  While both the 3D map and the {\it Planck} map detect the same dust, it is challenging to separate out the Orion dust ring from the background dust in the Crossbones and in the Galactic plane without the 3D PS1 map.  In the lower panel, the red, green, and blue rings approximately overlay three large ring-like structures in the region: Barnard's Loop, the new Orion dust ring, and the $\lambda$~Orionis dust ring.
}
\end{figure}

The projection of material in these different slices along the line of sight in far-infrared-based 2D dust maps renders separation of these different structures extremely challenging.  Figure~\ref{fig:orionplanck} shows our 3-color composite map, followed by the {\it Planck} far-infrared thermal dust map in the vicinity \citep{Planck:2013}.  While the Planck map makes clear that the dust we are tracing is real, separating the dust in the Orion dust ring from dust in the Crossbones or in the Galactic plane is not possible with the far-infrared data alone.

\subsection{Distance}
\label{subsec:distance}

\begin{figure*}[htb]
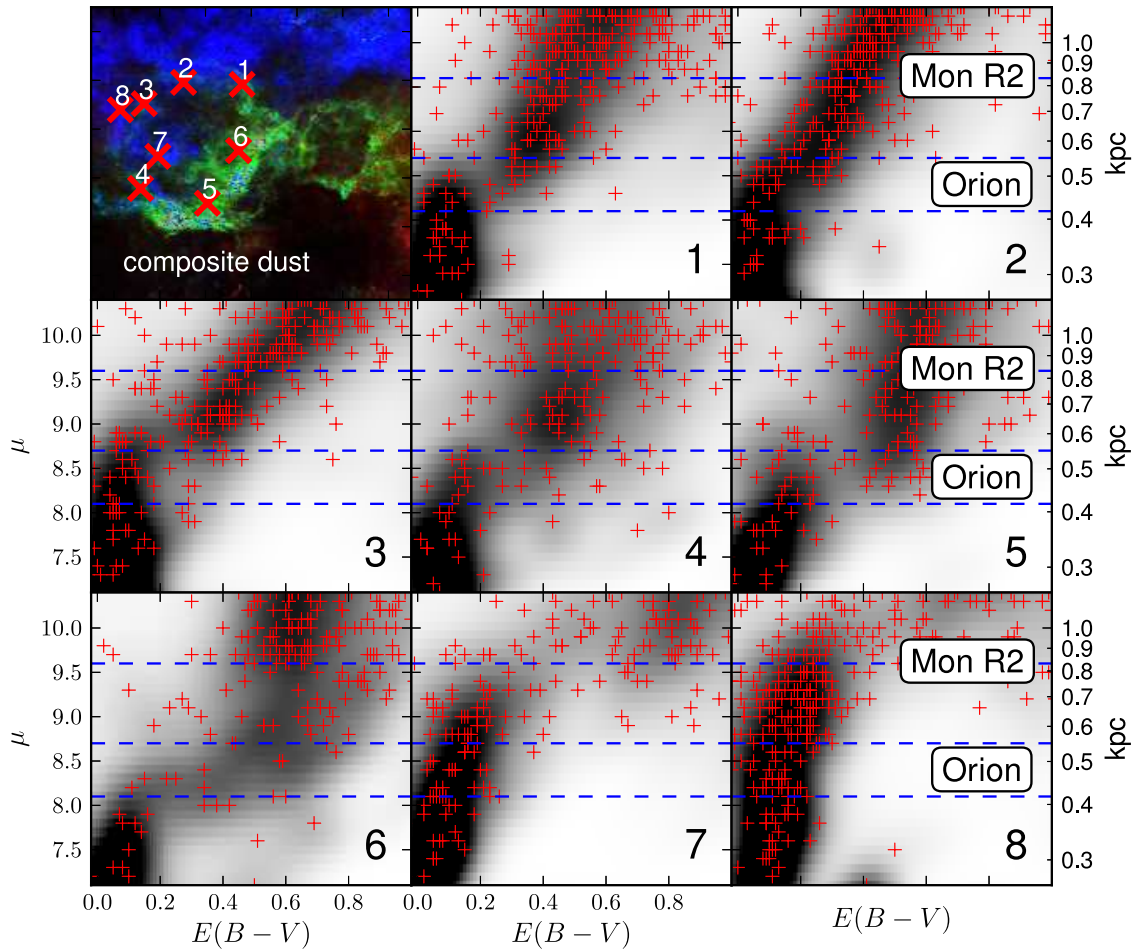

\dfplot{oriondistancesummary}
\figcaption{
\label{fig:oriondistance}
Distances and reddenings to stars along eight lines of sight in the vicinity of the Orion Molecular Complex.  The first panel shows our composite 3D map of the region, and gives the locations of the lines of sight by $\times$ symbols.  The lines of sight are labeled with numbers.  These numbers correspond to the numbers in the following eight panels, which show the expectation values of the distances (through the distance modulus $\mu = 5\log D / (10~\mathrm{pc})$) and reddenings $E(B-V)$ for stars within $0.2\degree$ of these sight lines (red crosses).  The background grayscales show the summed stellar posteriors.  The blue dashed horizontal lines correspond to the distances of the front of the ring at 414~pc, the back of the ring at 550~pc, and the distance to Monoceros R2 at 830~pc.  Sight lines 1--6 are chosen to pass through the Orion cloud and dust ring, and on these sight lines there is a clear increase in the reddenings of stars between 400 and 550~pc.  Sight lines 7 and 8 are chosen to pass through regions dominated by Monoceros R2, and show little sign of increased reddening between 400 and 550~pc, but significant increase in reddening at or beyond 830~pc, associated with Monoceros R2.
}
\end{figure*}

Our technique naturally allows the distance of the ring to be estimated.  The accuracy of this technique is limited to about 10\% by uncertainties in our stellar models and uncertainty in our priors on the distribution of stars and their metallicities in the Galaxy.  Moreover, the distance resolution is further limited by the small $7^\prime \times 7^\prime$ pixels in which we fit the 3D dust profiles.  We could obtain better estimates by using larger areas, as done in \citet{Schlafly:2014}.  However, the model of \citet{Schlafly:2014} assumes that the extinction is dominated by a single cloud, which is not true in the newly identified regions of most interest here.  Therefore, in order to get a handle on the distances to the different portions of the Orion dust ring, we show the expectation values of the distances and reddenings of stars in eight representative regions around the ring and in Monoceros R2 in Figure~\ref{fig:oriondistance}.  The first panel labels the lines of sight and shows their locations in Galactic latitude and longitude.  The following eight panels show the expectation values of the reddening (x-axis) and distance modulus (y-axis) for stars within $0.2\degree$ of the sight lines.  Horizontal dashed lines give the parallax distance to Orion of 414~pc from \citet{Menten:2007} (distance modulus $\mu = 8.08$), our distance estimate of the back edge of the ring of 550~pc ($\mu = 8.7$), and a standard distance to Mon~R2 of 830~pc \citet{Herbst:1976} ($\mu = 9.6$).  The work of \citet[L11]{Lombardi:2011} finds a somewhat larger distance ($905\pm37$~pc) to Mon~R2, while \citet{Schlafly:2014} find that various components of Mon~R2 reside from 830~pc to 1040~pc.  In any case, the line at 830~pc serves roughly to indicate the distance to Mon~R2.

The overall message of Figure~\ref{fig:oriondistance} is that lines of sight through the Orion dust ring show a sharp increase in reddening somewhere between 415 and 550~pc.  Given that the ring has a transverse diameter of about 100~pc, variation in line of sight distance at this level is not surprising.  That said, the lines of sight shown here are generally compatible with a distance of 480~pc; only line 6 is definitively more nearby, and only line 3 is definitively more distant.  This is broadly consistent with the results of \citet{Schlafly:2014}, which used the same techniques on a few lines of sight through Orion, and found a larger distance ($\approx 450$~pc) to the clouds on most of those lines, while also finding a distance consistent with the parallax distance of \citet{Menten:2007} along the lines closest to the Orion Nebula.

In order to try to get a better handle on possible distance variations to different components of the ring, we have repeated our analysis at lower angular resolution ($14^\prime$) and higher distance resolution ($\Delta \mu = 0.27$).  At this resolution the narrow filaments become poorly resolved, but the map of the region is qualitatively the same.  We then ``unwrap'' the ring around its center, and study the variation in distance to components of the ring as a function of angle.  To do this, we first compute, for each pixel, the distance from the center of the ring at ($l$, $b$) = ($212\degree$, $-11.5\degree$) and the angle clockwise around the center of the ring from Galactic North.  We select all pixels between 4.5\degree\ and 9.5\degree\ from the ring center.  We then select the subset of these pixels for which the reddening at the distance to Orion is more than 0.1 mag $E(B-V)$, after smoothing the reddening map with a Gaussian of 0.5\degree\ FWHM.  This selection is intended to limit the analysis only to pixels which show significant extinction at the distance to Orion.

\begin{figure*}[htb]
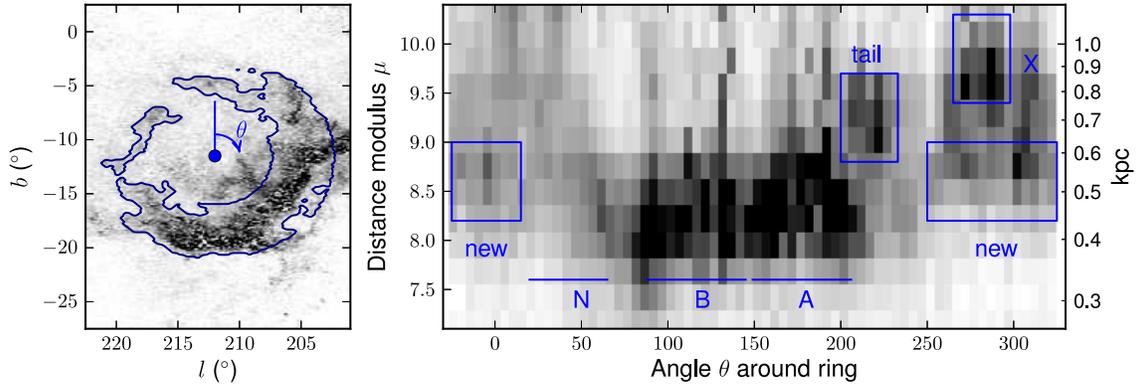

\dfplot{orionunwrap}
\figcaption{
\label{fig:orionunwrap}
Distribution of dust as a function of distance and of angle around the Orion dust ring, in degrees clockwise from Galactic North.  The left panel shows the pixels used in this analysis, circled in blue, on top of our estimated $E(B-V)$ at the distance to Orion (reproduced from Figure~\ref{fig:oriondust}).  The right panel shows the average amount of reddening for pixels in each angle and distance bin.  Angles corresponding to Orion A (A), Orion B (B), and the Northern Filament are marked, as are the new filaments containing the Orion dust ring (new), the Crossbones (X), and the ``tail'' of dust from the Crossbones that overlaps Orion A (tail).  The gray scale corresponds to 0--0.7~mag~$E(B-V)$ in the left panel, and 0--0.1~mag~$E(B-V)$ in the right panel.
}
\end{figure*}

We show these pixels and their average reddening at each distance and angle in Figure~\ref{fig:orionunwrap}.  Though we use only pixels with significant extinction at the distance to Orion, in the new filaments even these pixels have dramatically more dust in the background.  This, coupled with uncertainty in our distance estimates, makes a band of reddening at the distance to Orion difficult to discern at angles $\theta < 50\degree$ and $\theta > 250\degree$.  The morphological separation permitted by Figure~\ref{fig:oriondust} much more convincingly distinguishes the ring from background clouds.

The main molecular clouds Orion A and Orion B appear prominently in Figure~\ref{fig:orionunwrap} as a dark band at a distance of about 440~pc, $70\degree < \theta < 250\degree$.  The Northern Filament is also apparent, until the appearance of significant dust in its background at around $\theta = 40\degree$.  Clouds in the Crossbones are evident in the background of our new filament at $250\degree < \theta < 320\degree$, and are somewhat separated in distance from the new filament.  Likewise, the new filament in the foreground of the Galactic plane ($-25\degree < \theta < 15\degree$) is visible as increased reddening at Orion's distance, but at these angles there is also significant dust in the background.

In Figure~\ref{fig:orionunwrap} we find that Orion A, B, and the Northern Filament all share a common distance, $D = 440$~pc.  The end of the Northern Filament nearest the Galactic center is further away, with $D= 500$~pc, as is the new filament in the foreground of the Galactic plane ($-25\degree < \theta < 15\degree$).  The new filament in the foreground of the Crossbones may be the furthest away, with $D \approx 580~\mathrm{pc}$.  This is in good agreement with Figure~\ref{fig:oriondistance}, where line of sight 3, passing through this cloud, was found to have the most distant dust.  We caution that our absolute distance scale is uncertain at the 10\% level \citep{Schlafly:2014}, and that our pixels here have a resolution of only 15\% in distance, which likewise limits our accuracy.  Still, we believe the 100~pc variation in distance is real.

The final feature of Figure~\ref{fig:orionunwrap} worth mentioning is that at $\theta = 215\degree$ there is a prominent feature (marked ``tail'') in the background of Orion A, at a distance of approximately 680~pc.  This feature corresponds to the eastern tail of Orion A near ($l$, $b$) = (218.5\degree, -17.8\degree) and the filament linking the Crossbones and Orion A in projection near ($l$, $b$) = (215.9\degree, -15.8\degree).  Our analysis places these features behind Orion A, and somewhat in the foreground of the Crossbones.  The filament linking the Crossbones and Orion A also has a velocity significantly discrepant from the typical velocity in Orion A, which is more consistent with Mon R2; see \textsection\ref{subsec:vel}.

\subsection{Mass}
\label{subsec:mass}

We can make a rough estimate of the mass of the Orion dust ring by converting the observed column $E(B-V)$ to mass, which is dominated by atomic and molecular hydrogen.  We adopt the traditional ratio 
\begin{equation*}
\beta = (N(\HI) + 2N(H_2)) / E(B-V)
\end{equation*}
of $5.8 \times 10^{21}~\mathrm{cm}^{-2}~\mathrm{mag}^{-1}$ \citep{Bohlin:1978} for the hydrogen column density per magnitude of reddening.  The work of L11 considers the masses of various clouds in the vicinity of Orion via their extinction $A_K$; we largely follow their notation here.  The total mass $M$ is given by
\begin{equation}
M = D^2 \mu \beta \int E(\mathbf{\Omega}) d\mathbf{\Omega}
\end{equation}
where $D$ is the cloud distance, $\mu$ is the mean molecular weight, $\beta$ is the hydrogen column density to extinction ratio, and $E(\mathbf{\Omega})$ is the reddening $E(B-V)$ in the direction $\mathbf{\Omega}$.  We adopt $\mu = 1.37$ and $\beta$ from L11, in order to make our estimates as directly comparable to theirs as possible, converting from extinction $A_K$ to $E(B-V)$ using the results of \citet{Schlafly:2011}.  We consider only dust with $300 < D < 640~\mathrm{pc}$ as relevant, and treat all of this dust as if it were a thin screen at $D = 414$~pc \citep{Menten:2007}.  We follow L11 and define the following lines as boundaries between the different clouds in Orion:
\begin{align*}
  & \text{Orion A:} &             203^\circ \le l & {} \le 217^\circ \; , &   -21^\circ \le b & {} \le -17^\circ \; , \notag\\
  & \text{Orion B:} &             201^\circ \le l & {} \le 210^\circ \; , &   -17^\circ \le b & {} \le  -5^\circ \; , \notag\\
  & \text{$\lambda$~Orionis:} &   188^\circ \le l & {} \le 201^\circ \; , &   -18^\circ \le b & {} \le  -7^\circ \; . \notag\\
\end{align*}
We additionally define the entire Orion dust ring region and some filaments within it:
\begin{align*}
  & \text{Ring All:} &               201^\circ \le l & {} \le 222.5^\circ \; , &   -27.5 ^\circ \le b & {} \le   2.5^\circ \; , \notag\\ 
  & \text{Ring Plane:} &             210^\circ \le l & {} \le 215  ^\circ \; , &   -5.7  ^\circ \le b & {} \le  -3.5^\circ \; , \notag\\
  & \text{Ring East:} &              214^\circ \le l & {} \le 219  ^\circ \; , &   -10   ^\circ \le b & {} \le  -6.5^\circ \; . \notag\\
\end{align*}

Table~\ref{tab:mass} shows the masses we obtain within these regions as compared with L11.  We have rescaled the masses given in L11 to a fixed distance of 414~pc, because the work of \citet{Schlafly:2014} does not confirm a larger distance to $\lambda$~Orionis than to Orion A and B, as found in L11.  Our mass estimates are uniformly lower ($\sim 40\%$) than the L11 estimates, and are likewise lower than the similar estimates of \citet{Wilson:2005} and \citet{Ackerman:2012}.

We expect our mass estimates to be lower than those of other works for two reasons: first, we are intentionally excluding foreground and background dust from the region; we include only dust at the approximate distance of Orion.  In the case of Orion A, however, little foreground or background dust is expected to be present, yet we nevertheless underestimate the mass.  This is because uncertainty in our distance estimates leads some dust actually in Orion to be placed in front of or behind the cloud.  The problem is particularly severe in dense regions where there may be substantial variation in dust column within each pixel of our maps; in these cases, our technique tends to place substantial dust at large distances, which then is not counted toward the total mass.  Additionally, our technique saturates at an $E(B-V)$ of about 1.5~mag; the reddening map of \citet{Kainulainen:2009} indicates that only about 70\% of the mass of Orion A resides in clouds with $E(B-V) < 1.5$.  

Still, we mitigate these problems by tabulating in Table~\ref{tab:mass} also the total masses we would estimate, including all dust out to 2800~pc, but then assuming that this dust lies at 414~pc for the purposes of computation of the mass.  This procedure is more directly equivalent to the computation that L11 perform.  We note, however, that these estimates mix dust at the distance to Orion with significant quantities of background dust in Orion B and foreground dust in $\lambda$~Orionis; only 3D techniques like ours can appropriately exclude this material.  

The agreement between our measurements and those of L11 is only marginally better, however, when considering all of the dust out to 2800~pc.  Because of the saturation of dense clouds in Orion A, we continue to underestimate the mass of Orion A relative to other works.  On the other hand, using these distances we now estimate larger masses than L11 in Orion B and in $\lambda$~Orionis, by about 40\%.  The major contributor to the difference in these regions may be a zero point offset between our reddening map and theirs.  These regions are large, and much of the regions contain little dust.  An offset of 0.1~mag in $E(B-V)$ (0.03~mag $A_K$) can make a 20\% difference in the total inferred mass; we believe this is comparable to the uncertainty in the zero point of the L11 map.

An additional significant uncertainty is the correct factor $\beta$ to employ.  We adopt the value $5.8 \times 10^{21}~\mathrm{cm}^{-2}~\mathrm{mag}^{-1}$ from \citet{Bohlin:1978}, which is the same as adopted by L11.  However, recent estimates for $\beta$ have differed from this value by up to 40\% \citep{Schlegel:1998, Peek:2013, Liszt:2014}, and $\beta$ also depends on $R_V$ \citep{Draine:2003}.  Relatedly, we are ultimately comparing a reddening map based on optical colors to one based on near-infrared colors.  We have adopted an $R_V=3.1$ reddening law to compare these two maps, which may be inappropriate in these dense regions.

Finally, the derived masses depend on the adopted distances.  We have used throughout a fixed distance of 414~pc to be consistent with \citet{Menten:2007}.  However our preferred value for the distance to clouds in Orion is closer to 440~pc, albeit with a 10\% uncertainty in the overall distance scale.  The most distant cloud in the Orion dust ring (Ring East in Table~\ref{tab:mass}) lies about 580~pc away.  Adopting nevertheless a distance of 414~pc for this cloud incurs a factor of two error in mass.  

In conclusion, given all of these uncertainties, we estimate roughly that these masses are accurate to a factor of two.  The uncertainties are large, but they at least provide some sense for the masses of Orion A and B relative to the masses of the new filaments.  Orion A and B are much more massive than the new filaments ($\sim 15\times$).  The new filaments have approximately 4000 solar masses of material in the selected regions, and there is some dust throughout the ring.

\begin{deluxetable}{cccccc}
\tablewidth{\columnwidth}
\tablecaption{Masses of Clouds in the Orion Molecular Complex
\label{tab:mass}
}
\tablehead{
Cloud & \multicolumn{5}{c}{Mass ($10^3~M_\odot$)} \\\cline{2-6}
 & 300--640 & $<300$ & 640--2800 & $<2800$ & L11
}
\startdata
             Orion A &    37.0 &    16.4 &    21.4 &    74.8 &    94.3 \\
             Orion B &    65.6 &    29.9 &    40.4 &   135.9 &   102.9 \\
   $\lambda$ Orionis &    55.8 &    46.9 &    19.0 &   121.7 &    88.3 \\
            Ring All &   163.2 &   106.4 &   329.7 &   599.4 &    *    \\
          Ring Plane &     3.2 &     1.6 &    13.6 &    18.4 &    *    \\
           Ring East &     4.4 &     2.8 &    11.6 &    18.8 &    *    \\
\enddata
\tablecomments{
Masses of clouds in Orion, compared with the same estimates from \citet[L11]{Lombardi:2011}.  We list the mass of dust at the distance to Orion (that is, dust between 300~pc and 640~pc), as well as the foreground ($<300$~pc), background (640--2800~pc), and total mass ($<2800$~pc), if we had assumed that all this dust was at 414~pc.  Our estimates for the dust at the distance to Orion are uniformly lower than the \citet{Lombardi:2011} estimates, though some part of this problem is confusion between dust in Orion and dust in its foreground or background.
}
\end{deluxetable}

\section{The Ring as a Bubble}
\label{sec:discussion}

We have uncovered a 14\degree\ ring of dust that includes the active star-forming regions in Orion, as well as other dust clouds in the Galactic plane and foreground to Monoceros R2.  The ring is remarkably circular, suggesting a bubble origin.  Our distance estimates to the clumps of dust that make up the edge of the ring suggest that all of the material is between approximately 400~pc and 550~pc, making the ring's 150~pc depth comparable to its 100~pc width.

\subsection{Velocity Structure}
\label{subsec:vel}

An expanding shell of material should show expansion signatures in its velocity.  Detailed \HI\ and CO gas observations of the region are available, so signs of expansion could be observable in the gas line-of-sight velocity measurements.  Unfortunately, these signs are difficult to detect.

\begin{figure*}[htb]
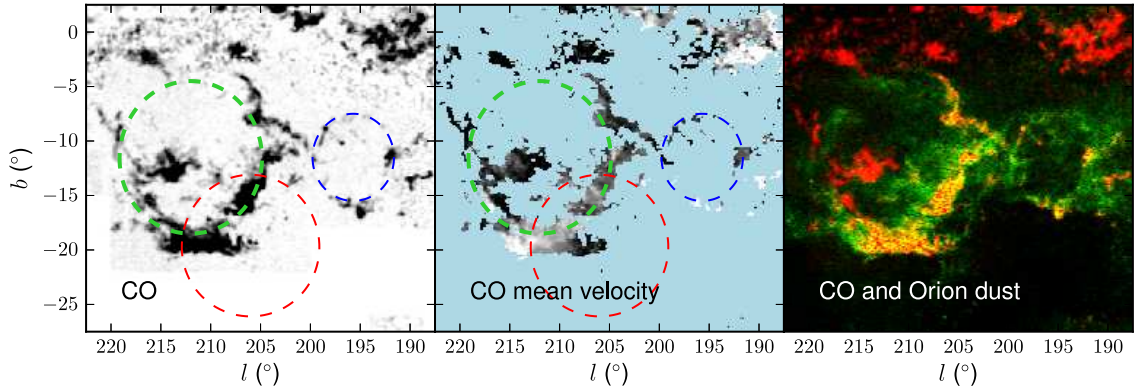

\dfplot{orionco}
\figcaption{
\label{fig:orionco}
The velocity-integrated CO intensity map of Orion from \citet{Dame:2001} (left panel). The northeast portion of the ring does not appear in CO, likely owing to an insufficient dust column to shield the molecular gas.  The intensity-weighted mean velocity in the velocity range 0--25~$\kms$ (center panel) shows a 5~\kms\ gradient from the northwest to southeast of the ring, and a clear velocity separation between the Monoceros R2 and Orion molecular gas (white to black spans 4--12~\kms).  In the third panel we show the CO intensity (red) and the Orion dust ring (green) on the same panel, to highlight the good agreement in the dense clouds but the absence of molecular material in the northeast.  The CO color scale corresponds to 0--13~$\mathrm{K~km~s}^{-1}$, and the $E(B-V)$ scale corresponds to 0--0.7~mag.
}
\end{figure*}

Figure~\ref{fig:orionco} shows the CO data from \citet{Dame:2001} and \citet{Wilson:2005} in the region of interest.  The first panel shows the total CO column in the area, with the Orion dust ring, the $\lambda$~Orionis ring, and Barnard's Loop overplotted.  The most prominent features in the Orion dust ring are clearly detected.  The new, formerly unrecognized filaments in the northeast of the ring, on the other hand, have no detectable CO, presumably due to their low dust column; they have $E(B-V) \approx 0.3$.  This prevents us from confirming the physical link between these filaments and Orion A and B via velocities.

The second panel of Figure~\ref{fig:orionco} shows the mean velocity of CO gas with $0 < v < 25~\kms$, with white to black spanning 4--12~\kms.  The most striking feature of the plot is the velocity difference between the Monoceros R2 and Orion clouds, though the different velocities are unsurprising given their different distances.  Our 3D maps split the eastern tail of Orion A (near ($l$, $b$) = (215\degree, -17\degree)) into two pieces: a piece at the distance to Orion, and a second piece that is nearer the distance to Mon~R2.  This split in distance is strikingly confirmed in velocity: the more distant component has the greater velocity characteristic of Monoceros R2, while the more nearby component has a velocity characteristic of Orion.  The work of \citet{Wilson:2005} considers this region as an expanding ring.  We confirm that the more distant part of their ring is moving away faster than the more nearby part, consistent with that picture.  However, we find that the separation between the front and back of their ring is about 300~pc, significantly larger than the 80~pc expected if the ring is circular.

As recognized by \citet{Wilson:2005}, there is a clear velocity gradient across the Orion Molecular Complex: the Northern Filament has a recession velocity about $5~\kms$ greater than Orion A.  If this motion is associated with the formation of the ring, it provides a very rough characteristic timescale of $100~\mathrm{pc}/(5~\kms) = 2\cdot10^7~\mathrm{yr}$, ignoring the geometry of the ring.  However, we do not favor this explanation for the velocity gradient.  The velocity gradient does not cleanly correlate with the variation in distance we observe, and the velocities of the clouds in Orion have been presumably heavily affected by winds and SN explosions in the Orion OB associations.

In principle, \HI\ observations should be sensitive to neutral gas in the northeastern dust clouds lacking CO.  However, there is neutral gas everywhere in the vicinity of Orion, and it is challenging to definitively correlate it with the clumps and filaments we observe in the dust.  Accordingly, we have focused on the simpler dust and CO measurements in this paper.

\subsection{\Ha\ emission in Orion}
\label{subsec:ha}

\begin{figure*}[htb]
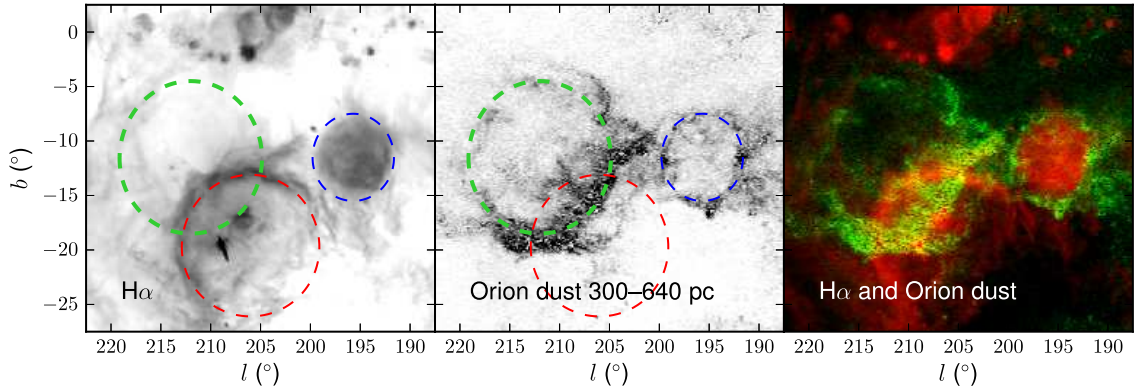

\dfplot{orionha}
\figcaption{
\label{fig:orionha}
\Ha\ map of Orion (left panel).  The three dashed circles show the location of the Orion dust ring, the $\lambda$~Orionis molecular ring, and Barnard's Loop.  A map of the dust at the distance to Orion is shown in the middle panel for comparison.  The right panel shows the \Ha\ data in red and the dust column in green, to show the tight correlation between the dust and \Ha\ in the $\lambda$~Orionis molecular ring, and its absence in the Orion dust ring.  The \Ha\ color scale is logarithmic from 10--1000~rayleigh in the left panel and from 10--315~rayleigh in the right panel.  The dust color scale ranges from 0--0.7~mag~$E(B-V)$ (linear).
}
\end{figure*}

Diffuse \Ha\ emission probes hot ionized hydrogen gas, such as that around young, high-mass stars, and is often characteristic of bubbles in the ISM.  Therefore we examine the Orion dust ring in \Ha\ in Figure~\ref{fig:orionha}, using the \Ha\ map of \citet{Finkbeiner:2003}, derived from the VTSS, SHASSA, and WHAM \Ha\ surveys \citep{Dennison:1999, Gaustad:2001, Reynolds:1998}.

There are two striking features visible in the \Ha\ map.  First, the $\lambda$~Orionis ring is filled with \Ha\ emission, just inside its associated dust ring.  This correlation has long been known; the \Ha\ emission is believed to be powered by ionizing emission from the O8 star $\lambda$~Orionis at the center of that ring, as discussed in \citet{Mathieu:2008}.

The second obvious feature is Barnard's Loop \citep{Barnard:1894}, the half-ring of \Ha\ emission centered near the Orion Nebula.  Though this \Ha\ loop has a similar size to the Orion dust ring, it is significantly offset angularly.  Given the close correlation between the \Ha\ and dust in $\lambda$~Orionis, it seems unlikely that Barnard's Loop and the Orion dust ring are physically linked.

\subsection{Emission at other frequencies}
Supernova remnants often feature X-ray emission \citep{Vink:2012}.  The Orion dust ring, however, shows no excess emission in the ROSAT all sky maps \citep{Voges:1999}.  Emission from the radioactive decay of $^{26}\mathrm{Al}$ can also reveal past supernovae.  There is a clear 1.8 MeV excess in the vicinity of Orion OB1a from COMPTEL \citep{Oberlack:1996}, but it is not well associated with the dust ring.

\subsection{The Orion Molecular Complex as a Bubble}
\label{subsec:bubble}

\begin{figure}[htb]
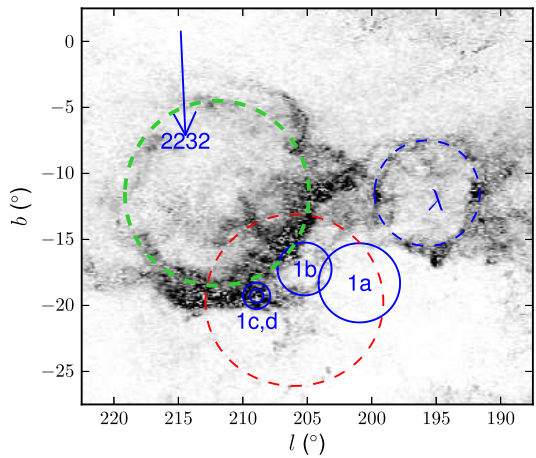

\dfplot{schematic}
\figcaption{
\label{fig:schematic}
Schematic map of open clusters and OB associations near Orion.  The open cluster NGC 2232 is marked by the number ``2232'', and its proper motion extrapolated over the last five million years is illustrated by the arrow.  The ``$\lambda$'' gives the location of $\lambda$~Orionis, and the solid circles give rough guides to the locations of various OB associations in Orion.  The three dashed circles show the Orion dust ring, the $\lambda$~Orionis ring, and Barnard's Loop.  The color scale ranges from 0--0.7~mag~$E(B-V)$.
}
\end{figure}

The circular ring of clouds in Orion strongly suggests a bubble origin, in which a shock wave swept up gas and dust which later collapsed, forming the observed molecular clouds \citep[e.g.,][]{Bally:2008b}.  One then immediately asks: what powered the bubble? How old is the bubble?  We have no clear answers, but Figure~\ref{fig:schematic} shows the bubbles in the context of some of the surrounding potential bubble sources in the region.

The locations of the Orion OB1 associations are indicated on Figure~\ref{fig:schematic} with solid circles.  Note that this Figure serves only to roughly show the extents of the associations; see \citet{Bally:2008b} for a detailed description of the complicated distribution of Orion OB1 stars.  These associations could certainly power a bubble.  However, the largest association, Orion OB1a, is believed to be the source of the Orion-Eridanus superbubble, and is outside the ring we observe.  It is hard to envision these stars producing both Barnard's Loop and the Orion dust ring given the large spatial offset between these two structures.  Given the energetic emission of the high-mass stars in Orion OB1a, Barnard's Loop and the Orion-Eridanus superbubble provide a much better match to Orion OB1a than the Orion dust ring does.  Meanwhile the younger Orion OB1 associations are embedded in Orion A, which seems unlikely if they formed a ring that includes Orion A.  Star formation in these regions may have instead been triggered by the bubble, or may have occurred naturally following the collapse of the molecular clouds.  Accordingly, it seems unlikely that the Orion OB1 associations formed the Orion dust ring.

The open cluster NGC 2232 is a plausible candidate.  The cluster is located at $(l, b) = (214.4\degree, -7.5\degree)$, within the projected interior of the bubble, and has an estimated parallax distance of 350~pc, with an uncertainty of about 7\% \citep{vanLeeuwen:2009}.  This distance places the cluster slightly in the foreground of the bubble, though a position within the bubble is not completely excluded.

The cluster has a supersolar metallicity of $\mathrm{[Fe/H]} \approx 0.25$ \citep{Monroe:2010}, and the work of \citet{Lyra:2006} estimates an age for this cluster of between 25 and 35 million years with uncertainties of about 25\%.  The proper motion of the cluster is $(\mu_l, \mu_b) = (-0.3, -6.2)~\mathrm{mas~yr}^{-1}$ out of the Galactic plane.  If the Orion dust ring is associated with NGC 2232, we would expect the gas and dust to have a similar large proper motion; however, the bulk proper motion of the gas and dust in the region is unknown.  The extrapolated proper motion of NGC 2232 over the last five million years is shown by the blue arrow on Figure~\ref{fig:schematic}.

How old is the dust ring?  Given the absence of any \Ha\ emission, the ring must be older than typical O star main sequence life times of about $10^7$ yr.  Such an age is also necessary to make the Orion dust ring predate the young OB associations like Orion OB1a that formed in Orion A.

The similarity between the Orion dust ring and the $\lambda$~Orionis ring encourages comparison.  The star $\lambda$~Orionis has an age of about five million years \citep{Mathieu:2008}, and the associated ring has a radius of about 8\degree\ at the same distance as the Orion dust ring.  During the wind-driven expansion of an adiabatic bubble, the radius of the bubble increases with $t^{3/5}$; in later stages, after the bubble becomes radiative or pressure confined, it grows less quickly with time \citep{Koo:1992}.  Since the Orion dust ring is twice as large as the $\lambda$~Orionis molecular ring, we would expect an age at least a factor of $2^{5/3}$ greater, or $t > 15$ million years.  However, the clear \Ha\ bubble in $\lambda$~Orionis suggests that expansion there is proceeding.  Meanwhile, the expansion of the Orion dust ring may have been halted when hot gas interior to the ring escaped and the bubble wall collapsed, so that the size of the ring is no longer a good proxy for its age.  Thus this estimate for the age of the Orion dust ring is a lower bound.  This model is complicated by the fact that \citet{Dolan:2002} and \citet{Cunha:1996} prefer a scenario in which the $\lambda$~Orionis molecular ring was formed by a supernova one million years ago, rather than the slower wind formation proposed by \citet{Maddalena:1987}.

A possible picture may be that the Orion bubble is a remnant of a $\lambda$~Orionis-like \HII\ bubble.  We envision a ring that once looked much like $\lambda$~Orionis, but after the sources powering the bubble have been exhausted and the hot gas has escaped.  The dust and collapsed molecular clouds trace the locations of the former bubble walls.  The gas velocity has since been influenced by subsequent star formation, leaving the dust ring and molecular gas as the remaining detectable signatures of the bubble.  The work of \citet{McKee:1984} considers photoionized bubbles in a clumpy medium, and finds that they have a characteristic homogenized radius of $56 n^{-0.3}~\mathrm{pc}$ after the ionizing star dies.  Here $n$ is the average number density of hydrogen in $\mathrm{cm}^{-3}$ over the entire sphere, which in this case is $M/(\mu V) \approx 11~\mathrm{cm}^{-3}$, using $200 \cdot 10^3 M_\odot$ as the total mass of the Orion Molecular Complex \citep{Wilson:2005}, a sphere of radius 50~pc for $V$, and $\mu = 1.37 \mathrm{~amu~atom}^{-1}$ as the mean molecular weight.  This leads to an estimate of homogenized radius of 27~pc; smaller than observed in the Orion dust ring, but of the same order of magnitude.  This discrepancy seems slight given that \citet{McKee:1984} consider models with a single star (though the dependence of the homogenization radius on the number of stars is weak).  

Typical ``superbubble'' models predict much larger radii than observed.  For example, \citet{deGeus:1992} considers bubbles created by stars in the Sco-Cen association, using the formulation of \citet{Weaver:1977} and \citet{McCray:1987}.  This gives
\begin{equation}
R = 269~\mathrm{pc}~(L_{38} / n_0)^{1/5} t_7^{3/5}
\end{equation}
with $R$ the radius of the bubble, $L_{38}$ the total mechanical luminosity of the stars in the association in units of $10^{38}~\mathrm{erg}~\mathrm{s}^{-1}$, and $t_7$ the time since the association was formed in units of $10^7~\mathrm{yr}$.  In order to get a radius as small as observed using an age of 20 million years, we need $L_{38} / n_0 = 2.8\times 10^{-5}$.  For our rough estimate of $n_0 = 11$, we required an extremely weak association ($L_{38} = 0.0003$), weaker than an individual O star.  However, it is possible that the radius is smaller than expected because the bubble expansion was halted when the bubble wall collapsed into the discrete clouds currently populating the ring, allowing the hot gas to escape.

Similarly, the observed bubble may also be smaller than expected if the bubble has ``blown out'' along the line of sight.  In a study of bubble features in the ISM, \citet{Beaumont:2010} find that many have no detectable gas or dust near the center of the bubble.  This is consistent with a picture where the natal cloud complex was sheet-like, and the observed rings appear as holes blown in the initial sheet.  The Orion dust ring does contain significant dust near the center of the ring, though most of the material is concentrated in the ring.

A final possibility is that the ring was formed by a supernova explosion.  The late stage behavior of SN remnants is considered in \citet{Cioffi:1988}.  They predict that a supernova remnant will merge with the interstellar medium at a size of about 20~pc for a density $n_0 = 11$, smaller than the Orion dust ring.  To obtain a 50~pc radius, one needs $n_0 \approx 1$.  Substantial numbers of SN remnants have been found with comparable sizes to the Orion dust ring \citep[e.g.,][]{Badenes:2010}, though these are presumed to come from SN in lower density environments.

\subsection{Alternative Interpretations}
\label{subsec:alternatives}

Our interpretation of the dust ring rests entirely upon the circularity of the ring.  We cannot rule out a chance alignment of clouds in a circular arrangement.  

Still, the only comparably circular, nearby, high-latitude dust ring we know of is the $\lambda$~Orionis molecular ring, which the \Ha\ data confirm to be of bubble origin.  Additionally, the Orion dust ring and the $\lambda$~Orionis dust ring contain nearly all of the dust in this volume of space.  There is a marked absence of dust, for instance, outside of the two rings in this region and at this distance, for example, at $b > -5\degree$.  This makes a chance alignment of various dust clouds unlikely, as these clouds are the only clouds in the region.

A second possibility is that the region is more complicated than we have described.  The primary accomplishment of our method in this region is to separate the Orion dust from background dust in Monoceros R2 and elsewhere in the Galactic plane.  This does reveal, however, a few features that are unsettling given the expected physical independence of Monoceros R2 and Orion, complicating the interpretation that Orion and Monoceros R2 are two unrelated clouds.  

The most interesting of these features lies in the region within 5\degree\ of ($l$, $b$) = (216\degree, -17\degree) (see Figure~\ref{fig:oriondust}).  In this region, there are a number of different structures.  There are two filaments, both detected in CO at different velocities (Figure~\ref{fig:orionco}), extending northeast from the edge of Orion A toward the Crossbones.  The different velocities and different distances we find for these filaments argues that these are simply chance projections, though \citet{Wilson:2005} considers the possibility they arise from an expanding ring.  A third, cometary filament extends east from Orion A in this region as well.  We find that this filament has a distance intermediate between Orion A and Monoceros R2 ($\approx 700~\mathrm{pc}$).  It also has a velocity larger than typical in Orion A.

Relatedly, a filament in the Orion dust ring and one in the Crossbones overlap in projection near ($l$, $b$) = (218\degree, -11\degree).  These filaments have surprisingly similar orientations and lengths.

These features may indicate that Orion and Monoceros R2 together form some larger complex.  The morphology is complex and suggests no simple explanation.  In general the good agreement in cloud distances around the entire Orion dust ring (Figure~\ref{fig:oriondistance}) argues for the simpler picture we have presented in this work.

\section{Conclusion}
\label{sec:conclusion}

We present 3D maps of dust reddening toward the Orion Molecular Complex.  The maps not only trace the total dust column also seen by {\it Planck}, but reveal additionally the distances to these clouds with unprecedented resolution.  This distance resolution allows clear separation of clouds at the distance of Orion from more distant clouds in the Monoceros R2 complex and further away in the Galactic plane.

Our maps of the dust at the distance of Orion unveil the ``Orion dust ring,'' a 100~pc ring of material that includes the main Orion A and B clouds.  The circular geometry of the ring and the close morphological similarity to the adjacent $\lambda$~Orionis ring motivates interpretation of the ring as a remnant of a bubble in the ISM.  Because we cannot identify the energy source of the bubble, we cannot classify the Orion dust ring as a remnant of an \HII\ region or a weak superbubble.  Still, the Orion Molecular Complex is among the best modeled and understood molecular clouds in the Milky Way \citep[e.g.,][]{Bally:2008b, Pon:2014}, and our results shed new light on its history and large scale structure.

The Orion dust ring may have implications for triggered star formation in the Orion Molecular Complex \citep[e.g.,][]{Lee:2009}.  Formation on a bubble wall may provide an explanation for why the Orion Molecular Cloud hosts dramatically more star formation than the California Molecular Cloud, despite their similar masses \citep{Harvey:2013, Lada:2009}.  Neveretheless, this is far from the whole story, as even within the Orion dust ring, Orion A and Orion B have very different star formation efficiencies \citep{Stutz:2013}.

The Orion Molecular Complex is an interesting test bed for 3D dust mapping.  The molecular complex lies mostly at high latitudes, and is therefore only mildly confused with foreground and background dust.  A small portion of the overall structure is confused with other dust structures in the region, though this portion proves crucial to our interpretation of the complex.  In future work, we will apply our technique to regions more deeply embedded in the Galactic plane, where confusion becomes much more severe.  The photometry and parallaxes from the recently launched Gaia mission will enable much more accurate 3D dust maps, broadening the range of applications of this technique.

This effort is progress toward better understanding the 3D structure of the Orion Molecular Complex itself; we have tantalizing evidence of variations in the distances of different components of the Orion dust ring.  The forthcoming Gould's Belt Very Large Array Survey of Orion will measure parallaxes to a number of stars in the region, placing further constraints on the 3D structure of the complex \citep{Kounkel:2014}.

This work has been greatly facilitated by valuable discussions with J. Bally, W. P. Chen, C. Heiles, E. Mamajek, C. McKee, and A. Stutz.  ES acknowledges funding by Sonderforschungsbereich SFB 881 ``The Milky Way System'' (subproject A3) of the German Research Foundation (DFG).  DF acknowledges support of NASA grant NNX10AD69G.  GMG and DPF are partially supported by NSF grant AST-1312891.  N.F.M. gratefully acknowledges the CNRS for support through PICS project PICS06183.

The Pan-STARRS1 Surveys (PS1) have been made possible through contributions of the Institute for Astronomy, the University of Hawaii, the Pan-STARRS Project Office, the Max-Planck Society and its participating institutes, the Max Planck Institute for Astronomy, Heidelberg and the Max Planck Institute for Extraterrestrial Physics, Garching, The Johns Hopkins University, Durham University, the University of Edinburgh, Queen's University Belfast, the Harvard-Smithsonian Center for Astrophysics, the Las Cumbres Observatory Global Telescope Network Incorporated, the National Central University of Taiwan, the Space Telescope Science Institute, the National Aeronautics and Space Administration under Grant No. NNX08AR22G issued through the Planetary Science Division of the NASA Science Mission Directorate, the National Science Foundation under Grant No. AST-1238877, the University of Maryland, and Eotvos Lorand University (ELTE).

\bibliography{2dmap}

\end{document}